\documentclass[aps,prd,twocolumn,amsmath,groupedaddress]{revtex4}
\usepackage[dvips]{epsfig}
\usepackage{latexsym}
\begin{document}

\title{The point-charge self-energy in Lee-Wick Theories}
\author{F.A. Barone}
\email{fbarone@unifei.edu.br}
\author{ G. Flores-Hidalgo}
\email{gfloreshidalgo@unifei.edu.br}
\affiliation{Universidade Federal de Itajub\'a, IFQ, Av. BPS 1303, Pinheirinho, cep 37500-903,
Itajub\'a, MG, Brazil.}
\author{A.A Nogueira}
\email{nogueira@ift.unesp.br}
\affiliation{IFT - R. Dr. Bento Teobaldo Ferraz, 271,
 01140-070, S\~ao Paulo, SP, Brazil.}

\begin{abstract}
In this paper we study the ultraviolet and infrared behaviour of the self energy of a point-like charge in the vector and scalar Lee-Wick electrodynamics in a $d+1$ dimensional space time. It is shown that in the vector case, the self energy is strictly ultraviolet finite up to $d=3$ spatial dimensions, finite in the renormalized sense for any $d$ odd, infrared divergent for $d=2$ and ultraviolet divergent for $d>2$ even. On the other hand, in the scalar case, the self energy is striclty finite for $d\leq 3$, and finite, in the renormalized sense, for any $d$ odd.
\end{abstract}
\maketitle
\baselineskip=20pt

One of the most remarkable features of the so called Lee-Wick electrodynamics is the fact that this theory leads to a finite self energy for a point-like charge in $3+1$ dimensions \cite{Podolsky42,Podolsky44,Podolsky48,LW69,LW70}, what has important implications in the quantum context, mainly in what concerns the renormalizability of the theory \cite{GrinsteinPRD2008,gc}.
Theories of superior derivatives for the scalar field has been also considered in the literature, maily after the propose of the so called Lee-Wick Standard Model (LWSM) \cite{Grinstein2008,Espinosa2008,Underwood2009,CaronePRD2009,RizzoJHEP2008,RizzoJHEP2007,Schat2008,KraussPRD2008,CuzinattoIJMPA2011,AcciolyMPLA2010,AcciolyMPLA2011,Accioly2010, CaronePLB2008}.

Among other subjects concerning the Lee-Wick electrodynamics, in the work of reference \cite{BHN} the self energy of a point-like charge in an arbitrary number of spatial dimensions was discussed. The presented results were speculative, not conclusive and indicated that the point-like particle self-energy is divergent for space dimensions higher than $3$.
That is an important subject in the context of theories with higher dimensions, in what concerns models with superior derivatives, because some well known results of Lee-Wick theories, which are valid in $3+1$ dimensions, are no longer applicable when the space has not $3$ dime dimensions.  

In this paper we show that, by using dimensional regularization, the self-energy of a point-like particle is finite when the space has an odd dimension and diverges when the space has even dimension. We also consider the self energy of a point-like source for the Klein-Gordon-Lee-Wick field, where there is two mass parameters involved.

Let-us start with the Lee-Wick electrodynamics. It is described by the lagrangian density\cite{LW69,LW70}	
\begin{equation}
\label{defL}
{\cal L}_{A}=-\frac{1}{4}F_{\mu\nu}F^{\mu\nu}-\frac{1}{4m^{2}}F_{\mu\nu}\partial_{\alpha}{\partial}^{\alpha}F^{\mu\nu}-\frac{{(\partial_{\mu}A^{\mu})}^2}{2\xi}-J_{\mu}A^{\mu}\ ,
\end{equation}
where $J^{\mu}$ is the vector external source,
\begin{eqnarray}
F_{\mu\nu}&=&\partial_{\mu}A_{\nu}-\partial_{\nu}A_{\mu}
\end{eqnarray}
is the field strength, $A^{\mu}$ is the vector potential and $m$ is a parameter with mass dimension. The third term on the right hand side of (\ref{defL}) was introduced in order to fix the gauge and $\xi$ is a gauge fixing parameter. The corresponding propagator is \cite{BHN}
\begin{eqnarray}
\label{propagador}
D_{\mu\nu}(x,y)&=&\int \frac{{d^{d+1}p}}{(2\pi)^{d+1}}
\Biggl(\frac{1}{p^{2}-m^{2}}-\frac{1}{p^{2}}\Biggr)
\left\{{\eta}_{\mu\nu}-\frac{p_{\mu}p_{\nu}}{p^{2}}\right.
\nonumber\\
& &\left.\biggl[1+\xi\biggl(\frac{p^{2}}{m^{2}}-1\biggr) \biggr]\right\}e^{-ip(x-y)}.
\end{eqnarray}
The energy of the system due to the presence of the source is given by \cite{BHN,Zee,BH}
\begin{equation}
\label{Egeral}
E_{A}=\lim_{T\to\infty}\frac{1}{2T}\int d^{d+1}x\ d^{d+1}y\ J^{\mu}(x)D_{\mu\nu}(x,y)J^{\nu}(y)\ .
\end{equation}

Now we take the source of a point-like stationary charge $\lambda$ placed at position ${\bf a}$ in a $d+1$ space-time
\begin{equation}
\label{point}
J_\mu=\lambda\eta_{\mu0}\delta^{d}({\bf x}-{\bf a})\ ,
\end{equation}
where $\delta$ is the Dirac delta function in $d$ dimensions.

Replacing above expression in Eq. (\ref{Egeral}) and performing the integrals in $x^{0}$, $p^{0}$ and $y^{0}$, we have
\begin{eqnarray}
\label{add1}
E_{A}=\frac{1}{2}\lambda^2 \int \frac{d^d{\bf p}}{(2\pi)^d}\left[\frac{1}{{\bf p}^2}-\frac{1}{{\bf p}^2+m^2}\right]\cr\cr
=\frac{1}{2}\lambda^2 m^2 \int \frac{d^d{\bf p}}{(2\pi)^d}\frac{1}{{\bf p}^{2}({\bf p}^{2}-m^{2})}\ .
\end{eqnarray}
Integrating in $d-$dimensional spherical coordinates, we can write,
\begin{equation}
\label{add2}
E_{A}=\frac{\lambda^2m^2}{(4\pi)^{d/2}\Gamma(d/2)}\lim_{\Lambda\to\infty}\int_{0}^{\Lambda} dp \frac{p^{d-3}}{p^2+m^2}\;.
\end{equation}
where we used the fact that the integral in the angular variables gives $2\pi^{d/2}/\Gamma(d/2)$, with $\Gamma$ standing for the Euler Gamma function.

Before we calculate the self energy (\ref{add1}), let-us first make a power counting analysis in order to investigate the ultraviolet behavior of expression (\ref{add2}). In this case one can discard the term proportional to $m^2$ in the denominator of the integrand in Eq. (\ref{add2}) and integrate from a given finite value $P$. Then, we get for $d\neq 4$,
\begin{eqnarray}
\label{add3}
E_{A}&\sim&\frac{\lambda^2m^2}{(4\pi)^{d/2}\Gamma(d/2)}\frac{\Lambda^{d-4}-P^{d-4}}{d-4}\Bigg|_{\Lambda\to\infty}
\nonumber\\
&\sim&\frac{\lambda^2m^2}{(4\pi)^{d/2}\Gamma(d/2)}\frac{\Lambda^{d-4}}{d-4}\Bigg|_{\Lambda\to\infty},d\neq 4
\end{eqnarray}
For $d=4$ we integrate Eq. (\ref{add2}),
\begin{eqnarray}
\label{add4}
E_{A}&=&\frac{\lambda^2m^2}{2(4\pi)^{d/2}\Gamma(d/2)}\ln\Biggl(\frac{\Lambda^{2}}{m^{2}}+1\Biggr)\Bigg|_{\Lambda\to\infty}\nonumber\\
&\sim&\frac{\lambda^2m^2}{2(4\pi)^{d/2}\Gamma(d/2)}\ln\Biggl(\frac{\Lambda^{2}}{m^{2}}\Biggr)\Bigg|_{\Lambda\to\infty},d=4.
\end{eqnarray}
From Eqs. (\ref{add3}) and (\ref{add4}) we conclude that the self energy is ultraviolet finite for $d\leq 3$ and divergent for $d>3$. 
Also,  from Eq. (\ref{add2}) one can see directly that we have an infrared divergence for $d=2$.

We shall use dimensional regularization in order to regularize the integral in Eq. (\ref{add1}). Rewriting it as
\begin{equation}
E_{A}
=\left.\frac{1}{2}\lambda^2 m^2 \mu^{d-2\omega} \int \frac{d^{2\omega}{\bf p}}{(2\pi)^{2\omega}}\frac{1}{{\bf p}^{2}({\bf p}^{2}-m^{2})}
\right|_{2\omega\to d}\ ,
\end{equation}
where $\mu$ is a parameter with mass dimension, integrating in spherical coordinates and using the formula \cite{kaku}
\begin{equation}
\int_{0}^{\infty}dr\frac{r^{\beta}}{(r^{2}+C^{2})^{\alpha}}=\frac{\Gamma\Bigl(\frac{1+\beta}{2}\Bigr)\Gamma\Bigl(\alpha-\frac{(1+\beta)}{2}\Bigr)}{2(C^{2})^{\alpha-(1+\beta)/2}\Gamma(\alpha)},
\end{equation}
we obtain for the self-energy of a point charge,
\begin{equation}
\label{add5}
E_{A}=-\left.\frac{\lambda^2m^{2\omega-2} \mu^{d-2\omega}}{2^{2\omega+1}\pi^{\omega}}\frac{\Gamma(\omega-1)\Gamma(2-\omega)}{\Gamma(\omega)}
\right|_{2\omega\to d}\ .
\end{equation}

Above expression can be simplifyed by using the Gamma function property $\Gamma(z+1)=z\Gamma(z)$,
\begin{equation}
\label{Energiaeletromag}
E_{A}=-\left.\frac{\lambda^2m^{2\omega-2}\mu^{d-2\omega}}{2^{2\omega+1}\pi^{\omega}}\Gamma(1-\omega)\right|_{2\omega\to d}\ .
\end{equation}

From this expression one can see that the point-charge self-energy (\ref{Energiaeletromag}) diverges for any $d$ even, $d=2\omega=2,4,6,...$ and is finite for any value of $d$ odd, with the result
\begin{equation}
\label{enerab}
E_{A}=-\frac{\lambda^2m^{d-2}}{2^{d+1}\pi^{d/2}}\Gamma(1-\frac{d}{2})\,,~d={\rm odd}.
\end{equation}
For instance, we have the following values for $d=1,3,5$ and $d=7$, 
\begin{eqnarray}
E_{A}(d=1)=-\frac{\lambda^{2}}{4m}\ \ ,\ \ E_{A}(d=3)=\frac{\lambda^{2}m}{8\pi}\cr\cr
E_{A}(d=5)=-\frac{\lambda^{2}m^{3}}{48\pi^{2}}\ \ ,\ \ E_{A}(d=7)=\frac{\lambda^{2}m^{5}}{480\pi^{3}}\ .
\end{eqnarray}

In order to split the divergences in the self-energy (\ref{Energiaeletromag}) for even spatial dimensions, we make the substitution $2\omega=d-\epsilon$, where $\epsilon\to 0$, what leads to
\begin{equation}
\label{Ereg}
E_{A}=\lim_{\epsilon\to0}-\frac{\lambda^2m^{d-2}}{2^{d+1-\epsilon}\pi^{(d-\epsilon)/2}}\left(\frac{\mu}{m}\right)^{\epsilon}
\Gamma(1-\frac{d+\epsilon}2)\ .
\end{equation}
Using the expansions
\begin{equation}
\label{exp1}
\Biggl(\frac{\mu}{m}\Biggr)^{\epsilon}\cong 1+\epsilon\ln(\mu/m)+\epsilon^{2}[\ln(\mu/m)]^{2}+{\cal O}(\epsilon^{2})
\end{equation}
and \cite{PRlivro}
\begin{eqnarray}
\Gamma(-n+\delta)&\cong&\frac{(-1)^{n}}{n!}\Bigg[\frac{1}{\delta}+\Psi(n+1)+\frac{1}{2}\delta\Bigg(\frac{\pi^{2}}{3}
+\Psi^{2}(n+1)\nonumber\\
& &-\Psi'(n+1)\Bigg)\Bigg]+{\cal O}(\delta^{2})\  ,\  n=0,1,2,...
\label{exp2}
\end{eqnarray}
where $\Psi(s)=\frac{d}{ds}\ln\Big(\Gamma(s)\Big)$, $\Psi'(s)=\frac{d}{ds}\Psi(s)$ and $\Psi(n+1)=1+\frac{1}{2}+...+\frac{1}{n}-\gamma$, with $\gamma$ standing for the 
Euler-Mascheroni constant, we rewrite Eq. (\ref{Ereg}), only for even values of $d$, in the form
\begin{eqnarray}
E_{A}&=&\frac{(-1)^{d/2}\lambda^2m^{d-2}}{2^{d+1}\pi^{d/2}\Gamma(d/2)}\lim_{\epsilon\to 0}\Bigg(\frac{2}{\epsilon}+
\Psi(d/2)\nonumber\\
& &+2\ln(\mu/m)\Bigg)\ ,~d={\rm even},
\label{regular1}
\end{eqnarray}
which is explictly divergent.

Next we consider the Lee-Wick model for the scalar field $\phi$, whose lagrangian density is \cite{BHN}
\begin{equation}
\label{modeloescalar}
{\cal L}=\frac{1}{2}\partial_{\mu}\phi\partial^{\mu}\phi+\frac{1}{2}\partial_{\mu}\phi
\frac{\partial_{\gamma}\partial^{\gamma}}{m^{2}}\partial^{\mu}\phi-\frac{1}{2}M^{2}\phi^{2}+J\phi\,,
\end{equation}
with the corresponding propagator
\begin{equation}
\label{propescalar}
D(x,y)=\int\frac{d^{d+1}p}{(2\pi)^{d+1}}\frac{m^{2}}{p^{4}-m^{2}p^{2}+M^{2}m^{2}}\exp[-ip(x-y)]\ .
\end{equation}

From (\ref{propescalar}) one can show that this model exhibits two massive poles for momentum square \cite{BHN}, namely
\begin{equation}
\label{defm+-}
m_{\pm}^{2}=\frac{m^{2}}{2}\Biggl(1\pm\sqrt{1-\frac{4M^{2}}{m^{2}}}\Biggr).
\end{equation}
In order to avoid tachyonic modes, one must take the restriction
\begin{equation}
0\leq\frac{4M^2}{m^{2}}\leq1\ .
\end{equation}

If $M^{2}=0$, we have a theory similar to the one studied in the previous case, for the vector field, with one massive mode, with mass $m$, and a massless one. This case is very similar to the one studied previously and has no novel physical properties.

If $0<4M^{2}/m^{2}<1$ we have two field modes with different non vanishing masses, $m_{+}$ and $m_{-}$, both of them lower than $m$ and $M$. In this case the propagator can be rewritten in the form 
\begin{eqnarray}
D(x,y)&=&\int\frac{d^{d+1}p}{(2\pi)^{d+1}}\Biggl(\frac{1}{p^{2}-m_{+}^{2}}-\frac{1}{p^{2}-m_{-}^{2}}\Biggr)\nonumber\\
& &
\times\frac{1}{\sqrt{1-\frac{4M^{2}}{m^{2}}}}\exp[-ip(x-y)]\ .
\end{eqnarray}
Once $m>2M$, the self-energy of a point-charge, $J=\lambda\delta^{d}({\bf x}-{\bf a})$, is given by
\begin{eqnarray}
\label{sca1}
E_{\phi}&=&\frac{1}{2}\frac{\lambda^2}{\sqrt{1-\frac{4M^{2}}{m^{2}}}}\int \frac{d^d{\bf p}}{(2\pi)^d}\left[\frac{1}{{\bf p}^2+m_+^2}-\frac{1}{{\bf p}^2+m_-^2}\right]\cr\cr
&=&\frac{\lambda^2m^2}{2}\int \frac{d^d{\bf p}}{(2\pi)^d}\frac{1}{({\bf p}^2+m_+^2)({\bf p}^2+m_-^2)}\,.
\end{eqnarray}

From above expression we conclude, by a power counting analysis, that the self-energy is finite for $d\leq 3$ and ultraviolet divergent for $d>3$. Following a similar procedure we have done for the vector case, it is not dificult to 
show that the regularized self-energy is given by
\begin{eqnarray}
\label{zxc1}
E_{\phi}&=&\frac{\lambda^{2}}{2^{2\omega+1}\pi^{\omega}}\frac{
(
\mu_+^{d-2\omega}m_{+}^{2\omega-2}-\mu_-^{d-2\omega}m_{-}^{2\omega-2})}{\sqrt{1-\frac{4M^{2}}{m^{2}}}}\nonumber\\
& & \times\left. \Gamma(1-\omega)\right|_{2\omega\to d}\,,
~~m>2M,
\end{eqnarray}
where $\mu_+$ and $\mu_-$ are two arbitrary parameters with mass dimension, introduced for each one of the integrals in  the first line of Eq. (\ref{sca1}). It is not difficult to see that the above expression is finite for $d$ odd. In this case the self-energy reads
\begin{eqnarray}
\label{zxc1ad}
E_{\phi}&=&\frac{\lambda^{2}}{2^{d+1}\pi^{d/2}}\frac{
(m_{+}^{d-2}-m_{-}^{d-2})}{\sqrt{1-\frac{4M^{2}}{m^{2}}}}
\Gamma(1-d/2)\,,\nonumber\\
& &~~~~d={\rm odd},~~m>2M.
\end{eqnarray}
On the other hand, when $d=2$, the integral (\ref{sca1}) is finite and can be solved easily,
\begin{equation}
\label{finite}
 E_\phi=\frac{\lambda^2}{4\pi\sqrt{1-\frac{M^2}{m^2}}}\ln\frac{m_-}{m_+},~d=2,~m>2M.
\end{equation}

When $d>3$ and even, the divergence in the selfenergy  can be splitted by setting $2\omega=d-\epsilon$ in (\ref{zxc1}) and taking the limit $\epsilon\to 0$.
Using the expansions (\ref{exp1}) and (\ref{exp2}) in (\ref{zxc1}) we have
\begin{eqnarray}
E_{\phi}&=&\lim_{\epsilon\to0}\frac{(-1)^{d/2}\lambda^{2}}{2^{d+1}\pi^{d/2}\Gamma(d/2)}\frac{1}{\sqrt{1-\frac{4M^{2}}{m^{2}}}}\Bigg[\frac{2}{\epsilon}(m_{+}^{d-2}-m_{-}^{d-2})\nonumber\\
& &+m_{+}^{d-2}\ln\frac{\mu_{+}}{m_{+}} -m_{-}^{d-2}\ln\frac{\mu_{-}}{m_{+}}    +m_+^{d-2}  \Psi(d/2)\nonumber\\
& &
-m_-^{d-2}\Psi(d/2)\Bigg]\,,d>2, ~d={\rm even}, m> 2M,
\label{regular2}
\end{eqnarray}
which is divergent once $m>2M$, what implies $m_{+}>m_{-}$.

Finally we consider the case  $m=2M$. Then, the propagator (\ref{propescalar}) reduces to the form
\begin{equation}
\label{zxc2}
D(x,y)=\int\frac{d^{d+1}p}{(2\pi)^{d+1}}\frac{m^{2}}{(p^{2}-m^{2}/2)^2}\exp[-ip(x-y)]
\end{equation}
and the self energy of a point charge is given by
\begin{equation}
E_{\phi}=-\frac{\lambda^{2}}{2}\int\frac{d^{d}{\bf p}}{(2\pi)^{d}}\frac{m^{2}}{({\bf p}^{2}+m^{2}/2)^{2}},~m=2M.
\label{added1}
\end{equation}
Again, this integral is finite for $d=1,2,3$ and divergent for $d>3$.  Regularizing above integral with similar methods previously employed,
we find a finite result for any $d$ odd, 
\begin{equation}
E_\phi=-\frac{\lambda^{2}m^{d-2}}{2^{3d/2-1}\pi^{d/2}}\Gamma(2-d/2),~d={\rm odd}, m=2M.
\end{equation}
For even values of $d$, the divergent part of (\ref{added1}), can be splitted by dimensional regularization  as in the previous cases. It is obtained
\begin{eqnarray}
E_{\phi}&=&\lim_{\epsilon\to0}-\frac{\lambda^{2}m^{d-2}}{2^{3d/2-1}\pi^{d/2}}\Bigg[\frac{2}{\epsilon}+2\ln\frac{\mu}{m}+\Psi(d/2-1)\Bigg],
\nonumber\\
& & d>3, ~d={\rm even},~ m=2M.
\label{regular3}
\end{eqnarray}
As in previous cases, it is explicitly divergent.

At this point we would like to remark that our results could be extendd to the case of charges distributions along branes with an arbitrary number of dimensions \cite{BHN}. For instance, in $3$ spatial dimensions, a charged plate exhibits a finite self energy, while a uniform charge distribution along a straigth line has a divergent self energy.

In summary, in the present work we showed that in the Lee-Wick electrodynamics the self energy of a point-like charge is strictly ultraviolet finite for $d=1$ and $3$ spatial dimensions. Also, for $d>3$ and odd, it has been showed the finiteness of the self energy in the renormalized sense. For $d>3$ but even, the self-energy can not be renormalized, because we have no way to redefine parameters in order to render expression (\ref{regular1}) finite. When $d=2$ the self energy is infrared divergent and can not be renormalized too. For the Klein-Gordon-Lee-Wick theory, we have a similar situation, with the distinction that when $d=2$ the self energy is finite. From the regularized self-energy for any $d>3$ and even, Eq.  (\ref{regular2}) for $m>2M$ and (\ref{regular3}) for $m=2M$, we see
that there is no way to render such quantities finite by redefining charge or mass parameters.

\noindent
{\bf Acknowledgments}

\noindent
F.A Barone is very grateful to CNPq for partial financial support. G. Flores-Hidalgo thanks FAPEMIG por partial support. A.A. Nogueira thanks CAPES for financial support.



\end{document}